\documentclass{article}
\pdfoutput=1
\usepackage{hyperref}
\hypersetup{
  pdfinfo={
    Title={UNSUPERVISED ADVERSARIAL DOMAIN ADAPTATION FOR BARRETT’SSEGMENTATION},
    Author={Numan Celik, Soumya Gupta, Sharib Ali, Jens Rittscher},
  }
}

\usepackage{pdfpages}
\usepackage{spconf,amsmath,graphicx}
\usepackage[utf8]{inputenc}
\usepackage[english]{babel}
\usepackage{color}
\usepackage{amsfonts}
\usepackage{url}
\usepackage[mathscr]{euscript}
\let\euscr\mathscr \let\mathscr\relax
\usepackage[scr]{rsfso}
\usepackage{caption}
\usepackage{multirow}
\usepackage{tabularx,booktabs}
\newcolumntype{C}{>{\centering\arraybackslash}X} 
\usepackage{multicol}
\usepackage{algorithm2e}
\usepackage{booktabs,ragged2e}
\usepackage[utf8]{inputenc}
\usepackage{todonotes}

\usepackage{makecell}

\usepackage[flushleft]{threeparttable}
\usepackage{todonotes}
\usepackage{subcaption}

\def\x{{\mathbf x}}
\def\L{{\cal L}}

\begin{document}


\title{Unsupervised Adversarial Domain Adaptation for Barrett's segmentation}

\name{Numan Celik$^{1}$ \quad Soumya Gupta$^{1}$ \quad Sharib Ali$^{1,2}$ \quad  Jens Rittscher$^{1,2,3}$}

\address{$^{1}$ \small{Institute of Biomedical Engineering (IBME), Department of Engineering Science, University of Oxford, Oxford, UK} \\ $^{2}$ \small{NIHR Oxford Biomedical Research Centre, Oxford University Hospitals NHS Foundation Trust, Oxford, Oxfordshire, UK}\\ $^{3}$  \small{Nuffield Department of Surgical Sciences, University of Oxford, John Radcliffe Hospital, Oxford, UK}}
%
%
\maketitle
\def\x{{\mathbf x}}
\def\L{{\cal L}}
\begin{abstract}
Barrett's oesophagus (BE) is one of the early indicators of esophageal cancer. Patients with BE are monitored and undergo ablation therapies to minimise the risk, thereby making it eminent to identify the BE area precisely. Automated segmentation can help clinical endoscopists to assess and treat BE area more accurately. Endoscopy imaging of BE can include multiple modalities in addition to the conventional white light (WL) modality. Supervised models require large amount of manual annotations incorporating all data variability in the training data. However, it becomes cumbersome, tedious and labour intensive work to generate manual annotations, and additionally modality specific expertise is required. In this work, we aim to alleviate this problem by applying an unsupervised domain adaptation technique (UDA). Here, UDA is trained on white light endoscopy images as source domain and are well-adapted to generalise to produce segmentation on different imaging modalities as target domain, namely narrow band imaging and post acetic-acid WL imaging. Our dataset consists of a total of 871 images consisting of both source and target domains. Our results show that the UDA-based approach outperforms traditional supervised U-Net segmentation by nearly 10\% on both Dice similarity coefficient and intersection-over-union. 
\end{abstract}
\vspace{-0.3cm}
\section{Introduction}
\label{sec:intro}
Oesophageal adenocarcinoma (OAC) is the 7th most common cause of cancer deaths with a 5-years survival rate of only 14\%. One common precursor of OAC is Barrett's oesophagus (BE), which develops in about 10–20\% of patients with chronic gastroesophageal reflux disease or inflammation of the oesophagus~\cite{cook2018cancer}. If detected early, OAC can be treated, which is why early detection of OAC is critical to improve patient survival.~Endoscopy is therefore recommended for surveillance of patients with BE. Expert endoscopists can identify areas at risk of developing cancer, but this is challenging. Changes in the surface structure of the tissue and other irregularities indicate precancerous and cancer development. Identifying the extent of inflammation is an important measure of risk stratification. Here, automated segmentation can be a crucial step to delineate the region of interest (\textit{BE in this case}) more robustly and reliably.
\begin{figure}[t!]
\centering
\includegraphics[width=0.48\textwidth]{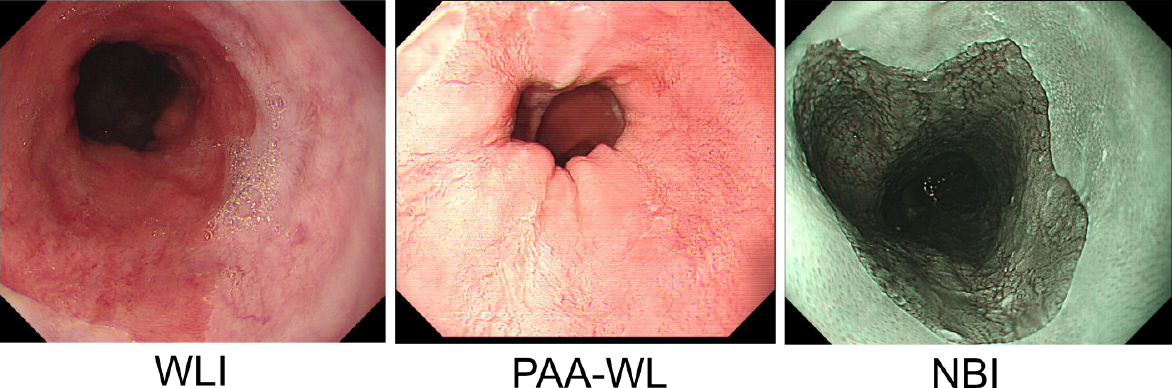}
\caption{Image modalities used in source (white light imaging, WLI) and target domain (post-acetic acid, PAA-WL and narrow band imaging, NBI).~\label{fig:1}}
\end{figure}
\begin{figure*}[t!]
  \centering
    \includegraphics[width=\textwidth]{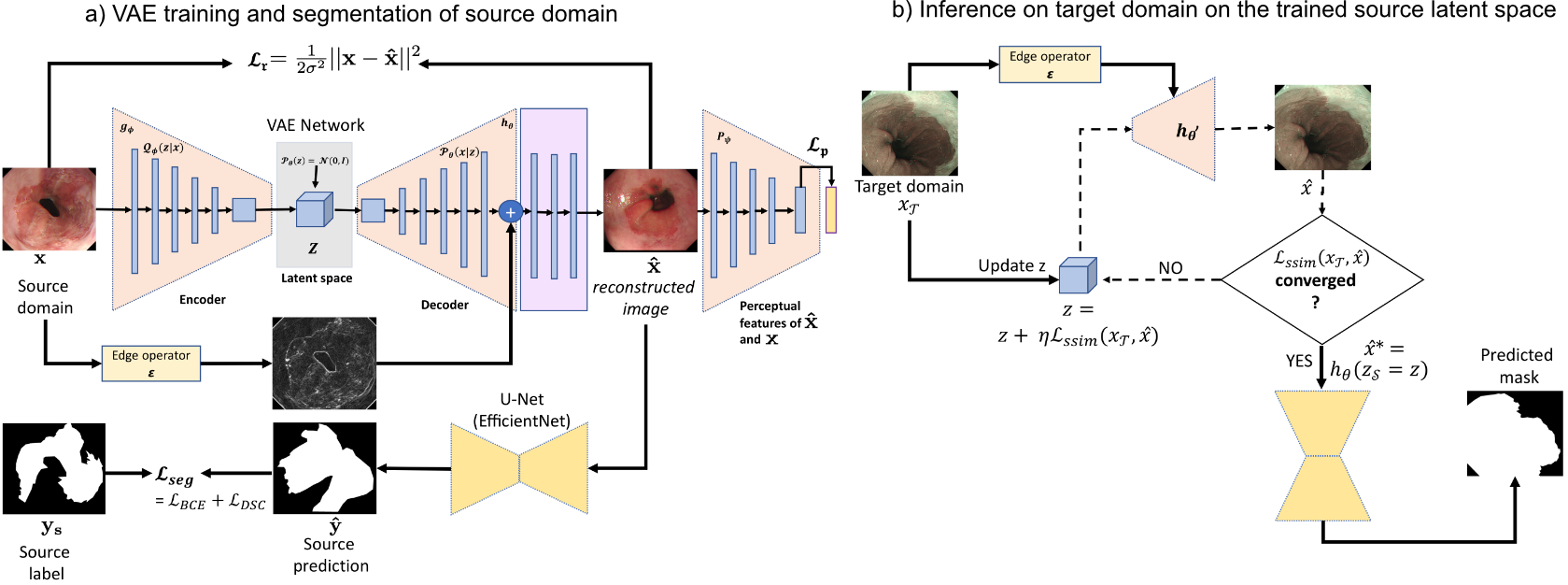}
    \caption{Complete architecture of adversarial unsupervised domain adaptation. a) Training of source domain where latent embedding is learnt with a reconstruction loss, KL-divergence loss and perceptual loss. In addition, the edge maps are added to the feature map of the decoder. U-Net is trained to segment the reconstructed image. b) Target domain images are mapped to the latent vector space and iterated to minimise structure similarity loss before input to the U-Net.~\label{fig:2}}
\end{figure*}
Convolutional Neural Network (CNN) based approaches such as Fully Convolutional Network (FCN)~\cite{FCN}, U-Net~\cite{Olaf:MICCAI2015}, DeepLab~\cite{DeepLabv3} are known to perform well in the biomedical image segmentation domain, including segmentation tasks on gastrointestinal (GI) endoscopy~\cite{ali2020objective,EndoCV2020,guo2020polyp,Ali20203D}. Here, the majority of teams also used U-Net-based models for multi-class GI disease segmentation challenge~\cite{EndoCV2020}. Gupta et al.~\cite{gupta2020mi} proposed a motion induced U-Net based segmentation strategy for kidney stone segmentation in the ureteroscopy data. Guo et al.~\cite{guo2020polyp} utilised a fully convolutional network based on atrous kernels to segment polyps. Similarly, DeepLabv3+ with ResNet50 backbone was used to segment BE from esophageal endoscopy data~\cite{Ali20203D}. 
However, these fully supervised models often do not generalise to different imaging modalities due to domain shift, and often obtaining reliable annotations for all these modalities require domain expertise and tremendous manual labeling time. Thus, to address this issue, domain adaptation~\cite{ADDA} and knowledge distillation~\cite{KDSS} approaches have been proposed for unsupervised or semi-supervised domain learning where annotations are not required in the target (test) dataset.

Tsai et al.~\cite{Tsai} developed a domain adaptation model for semantic segmentation based on a multi-level adversarial network which shares the segmentation loss for both source and target domains. In the discriminator network, target domain predictions are matched to the source domain predictions whether the input is from the source or target domain via an adversarial loss. In the context of biomedical image segmentation, Dong et al.~\cite{CardioUDA} proposed an unsupervised domain adaptation (UDA) technique for cardiothoracic ratio estimation by performing domain invariant chest organ segmentation. Haq et al.~\cite{haq20a} introduced an alternative UDA framework that included a decoder network to reconstruct the input images for cell segmentation. 

The existing GI endoscopy image segmentation methods are limited by the fully supervised approaches~\cite{EndoCV2020,guo2020polyp,Ali20203D}. To the best of our knowledge, this is the first report on segmentation of BE in endoscopy which utilises unsupervised domain adaptation. We use a target-independent approach which means that the network is only trained on the source domain. By adapting the UDA architecture~\cite{pandey2020unsupervised}, we design a semantic segmentation for the Barrett's area. This method involves a variational autoencoder (VAE) that produces a latent vector encoding of the source image and used to compute the nearest clone of the target image that is projected on this latent space. To do so, a structural similarity loss \(\euscr{L}_{ssim}\), and a U-Net~\cite{Olaf:MICCAI2015} model with a EfficientNet backbone is trained to predict the target image. The network is trained end-to-end with similar configuration. 
\begin{table*}[t!h!]
\caption{Comparison of the UDA framework for Barrett's segmentation (BarrettsSegUDA) with U-Net architecture.}
\centering
\begin{tabular}{lcccc|cccc}
    \toprule
    \multicolumn{1}{c|}{}& \multicolumn{4}{c|}{\textbf{WLI $\rightarrow$ NBI}}&\multicolumn{4}{c}{\textbf{WLI $\rightarrow$ post acetic acid}}\\
    \cline{2-9}
     \multicolumn{1}{c|}{\textbf{Method}}& \multicolumn{2}{c}{\textbf{Mixed}}&\multicolumn{2}{c|}{\textbf{Separated}}& \multicolumn{2}{c}{\textbf{Mixed}}&\multicolumn{2}{c}{\textbf{Separated}}\\
    \cline{1-9}
          & \multicolumn{1}{|c}{IoU} & DSC & IoU & DSC & IoU & DSC & IoU & DSC \\
    \cline{2-9}
     \multicolumn{1}{c|}{{U-Net}} & 0.765  & 0.885  & 0.657  & 0.791  & 0.803  & 0.916  & 0.717   & 0.828\\
    \cline{2-9}
     \multicolumn{1}{c|}{{BarrettsSegUDA}} & 0.767  & 0.881   & 0.762   & 0.879 & 0.798   & 0.901 & 0.774   & 0.893\\
    \bottomrule
\end{tabular}
\label{tab:comparision}
\end{table*}
Our experimental results demonstrate that  the proposed UDA method generalises on different imaging modalities showing improved segmentation accuracy when compared to the performance of the baseline U-Net architecture.
\vspace{-0.25cm}
\section{Method} 
The proposed Barrett's oesophagus (BE) segmentation approach is based on the UDA framework which consists of three main modules: (i) a VAE based generative model trained only on source domain (Fig.~\ref{fig:2} a), ii) a U-Net architecture for segmentation, and (iii) inference on target domain using VAE latent optimizing by optimising \(L_{ssim}\) (Fig.~\ref{fig:2} b). Below we will detail each component in detail.
%
%
\subsection{VAE training on source domain}
VAEs are the latent space derived generative models that are based on variational inference distributions where \begin{math}\mathcal{Q}_\phi (\textbf{z}\vert \textbf{x})\end{math} approximates the true posterior \begin{math}\euscr{P}_\theta (\textbf{z}\vert \textbf{x})\end{math}.
The Evidence Lower Bound term $\euscr{L}(\theta,\phi)$ in Eq.~(\ref{eq:1}) eliminates the non-negative KL-term making a lower bound on the data log-likelihood only which is maximized in VAE by parameterising $\euscr{Q}_\phi(\textbf{z}\vert \textbf{x})$ and $\euscr{P}_\theta (\textbf{x}\vert \textbf{z})$.  
\begin{equation}{\label{eq:1}}
\ln{(\euscr{P}_\theta (\textbf{x}))} = \euscr{L}(\theta,\phi) + \mathbb{D}_\textit{KL}[\euscr{Q}_\phi(\textbf{z}\vert \textbf{x})\|\euscr{P}_\theta (\textbf{z})]
\end{equation}

where, 
\begin{equation}
\euscr{L}(\theta,\phi) = \mathbb{E}_\euscr{Q_\phi}[\ln{(\euscr{P}_\theta (\textbf{x}\vert \textbf{z}))}] - \mathbb{D}_\textit{KL}[\euscr{Q}_{\phi}(\textbf{z}\vert \textbf{x})\|\euscr{P}_{\theta} (\textbf{z})] 
\end{equation}
Following the VAE framework \cite{pandey2020unsupervised}, we use two neural networks: an encoder network $g_\phi$ to compute the approximate posterior; and a decoder network $h_\theta$ to compute the image likelihood of the observed data \begin{math}\euscr{P}_\theta (\textbf{x}\vert \textbf{z})\end{math}. During training, the ``reconstructed'' \textbf{x} image is sampled in a two-step process: (i) sample \textbf{z} $\sim$ $\euscr{N}$(0,~\textit{I}) and (ii) sample \textbf{x} from \begin{math}\euscr{P}_\theta (\textbf{x}\vert \textbf{z})\end{math} at the outer layer of the decoder network. 

Together with the reconstruction loss $\euscr{L}_r$, a perceptual loss $\euscr{L}_p$ is also computed. While $\euscr{L}_r$ is computed as an $l2-$norm between the source sample and the reconstructed sample, perceptual loss is computed as an $l2-$norm between their feature embedding. Such a combined loss helps to obtain more sharper images compared to classical standard reconstruction loss alone. In addition, as shown in Fig.~\ref{fig:2} a), the edges of the input images are extracted and concatenated into the decoder network in VAE through a skip connection to better generalise the source and target domains. Fig.~\ref{fig:2}a) shows a schematic diagram of a complete VAE architecture used for training on the source domain. 
%
\subsection{Segmentation network}
The segmentation network consists of a U-Net where an EfficientNet-B4 is used as a backbone, wherein its input is the reconstructed image obtained from the decoder. A combined loss function consisting of dice loss $\euscr{L}_{DSC}$ and binary cross entropy loss $\euscr{L}_{BCE}$ is minimised between the predicted masks and the source ground truth masks. 
\subsection{Inference on target domain via VAE latent search}
The trained VAE encoding aims to find the nearest point from the source domain, given a sample from the target domain, through an optimisation over the latent space representation \textbf{z}, as can be seen in Fig.~\ref{fig:2} b. Once the decoder $h_\theta$ of the VAE is trained on the source distribution $\euscr{P}_s(\textbf{x})$, given an image $\tilde{\textbf{x}}_t$ from the target distribution, the latent search algorithm finds the nearest point by optimizing latent vector over the iterative process. As shown in the Fig.~\ref{fig:1}, the WL images are used in the source domain while the PAA and NBI are used in the target domain. The optimization procedure is performed by minimizing the structural similarity index (SSIM) based loss \(\euscr{L}_{ssim}\) between the image $\tilde{\textbf{x}}_t$ from target domain and reconstructed target image $\tilde{\textbf{x}}_s$ from the VAE. The SSIM for a pair of images (\textbf{x},$\hat{\textbf{x}})$ is defined as:
\begin{equation}
SSIM(\textbf{x},\hat{\textbf{x}}) = l(\textbf{x},\hat{\textbf{x}})^\alpha * c(\textbf{x},\hat{\textbf{x}})^\beta * s(\textbf{x},\hat{\textbf{x}})^\gamma
\end{equation}
Here \textit{l}, \textit{c} and \textit{s} denote luminance, contrast and structure similarities of the given pair of images, respectively. The parameters $\alpha$, $\beta$ and $\gamma$ $>$ 0; and the similarity loss function \(L_{ssim}\) is given by:
\begin{equation}
\euscr{L}_{ssim}(\textbf{x},\hat{\textbf{x}}) = 1 - SSIM(\textbf{x},\hat{\textbf{x}})
\end{equation}

After the convergence of \(\euscr{L}_{ssim}\) the optimal latent space is used to generate the closest clone of source domain which is then used as an input to the segmentation network to predict the target mask. 
%
\begin{figure*}[t!]
  \centering
    \includegraphics[width=0.9\textwidth]{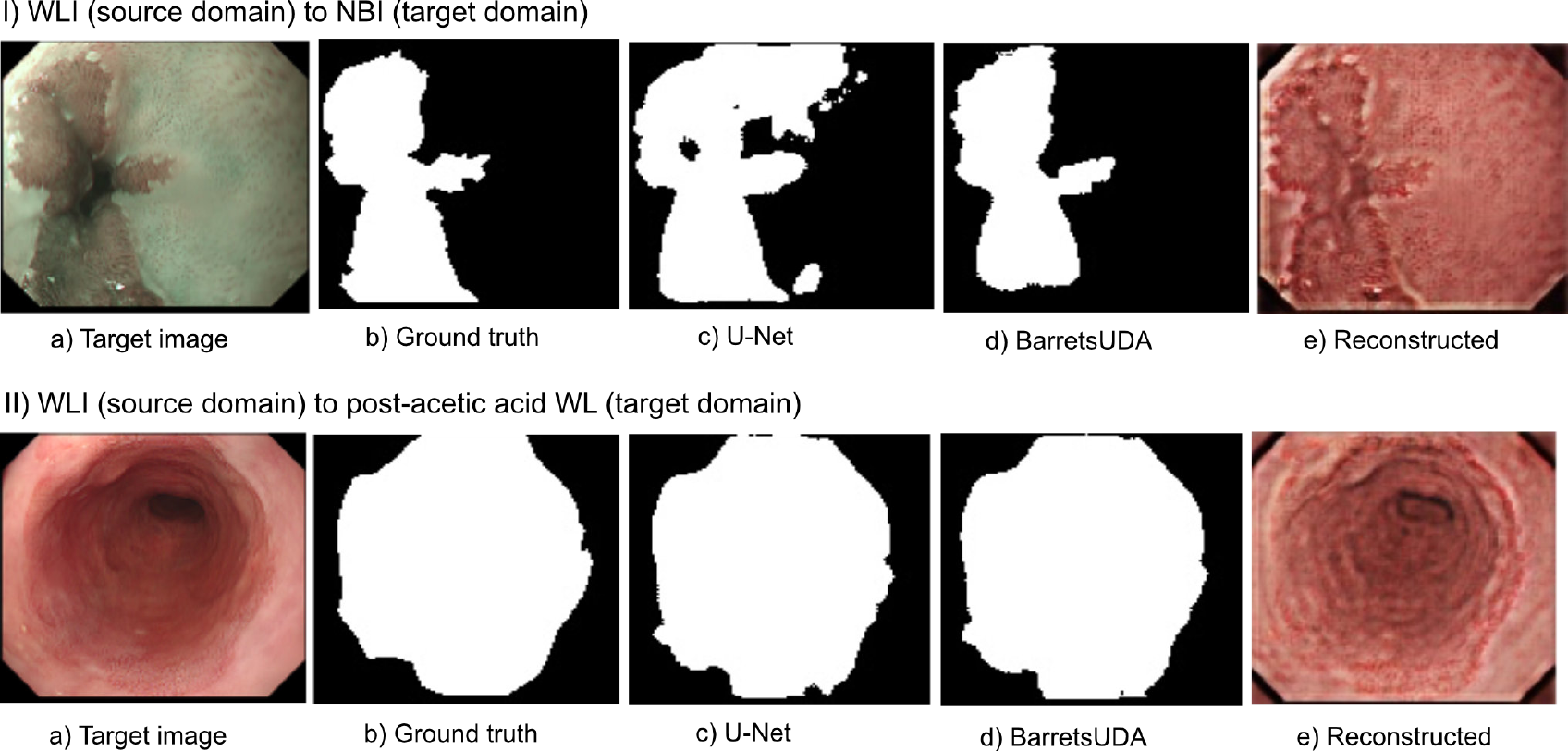}
    \caption{Visualization of segmentation of Barrett's oesophagus (BE) for both experiments. (a) and (b) indicate sample image and corresponding ground truth mask of the target domain. Segmentation of the given sample image by U-Net is shown in (c) and by the proposed UDA network in (d). The reconstructed image from VAE network is shown in (e).~\label{fig:qualitativeResults}}
\end{figure*}
\section{Experiments and results}
\subsection{{Dataset}}
We have curated a new dataset that consists of a total of 970 still video image frames acquired from nearly 100 multi-patient videos. All images represent the oesophageal area of patients with Barrett's oesophagus, a well-known precursor for oesophageal cancer. The videos were acquired during gastroesophageal endoscopy at the John Radcliffe Hospital, Oxford. All annotations were performed by two experienced post-doctoral researchers, one PhD student, and reviewed by two senior gastroenterologists.  

We evaluated the approach on the presented BE dataset by splitting it into three sub-datasets: WLI (source domain), NBI (target domain) and PAA (target domain) with 589 (80--20 train and validation split), 282 and 99 images, respectively. 
\subsection{Implementation Details}
%
The VAE architecture is shown in Fig.~\ref{fig:2} for both training and inference stage. In all experiments, the images were resized to $128 \times 128$ pixels. For VAE training process, the following loss functions are utilized: reconstruction loss $L_{r}$ as the mean squared error, and the perceptual loss is computed from the features extracted from the 6$\textsuperscript{th}$ layer of the segmentation network~\cite{tan2020efficientnet} and is trained on the source domain, WLI. An RMSProp with learning rate of 0.001 is used as an optimizer for segmentation network. The VAE network is trained for 100 epochs with a batch size of 32 with the same optimiser RMSprop but with a learning rate of 0.0001.
\subsection{Results}
%
We have evaluated our methods on NBI and post-acetic acid WL modalities (target domains) using two standard metrics: intersection-over union (IoU) and dice coefficient (DSC). Training for both these cases was done only on the WLI modality. However, for measuring the efficacy of the method, we also trained on mixed source image samples and tested on the mixed target samples with the shared 25\% data in each case. We also provide a comparison with the tradition U-Net model. 
\paragraph*{\textbf{WLI $\rightarrow$ NBI:}} In this case, we choose WLI as the source domain and NBI as the target domain (WLI $\rightarrow$ NBI). Table~\ref{tab:comparision} shows the empirical analysis of the applied UDA structure for BE segmentation against the standard U-Net~\cite{Olaf:MICCAI2015} model. We initially assessed the performance of the UDA model trained only on the source domain and tested on the target domain (referred as ``separated''). 
We further investigated the effect of adding some images from training domain into the target domain (referred as ``mixed'') on the segmentation. For the mixed, 25\% of WLI images were merged into the target domain test.  Table~\ref{tab:comparision} shows that, the BarrettsSegUDA gives 10.5\% higher IoU and 8.8\% better DSC compared to the traditional U-Net model in the target separated dataset. For the mixed target-source dataset, the proposed model output is comparable to the standard U-Net model. 
Fig.~\ref{fig:qualitativeResults} (I) shows that BarrettsSegUDA is more close to the ground truth mask compared to that of the standard U-Net model in segmentation, thereby indicating the effectiveness of the latent space optimisation algorithm in the inference module.

\paragraph*{\textbf{WLI $\rightarrow$ PAA images:}} In the second case, we conduct further domain generalisation experiment with the UDA approach using WLI as the source and post-acetic acid WL images as the target domain. Table~\ref{tab:comparision} compares the performance of the UDA model with the classical U-Net model along with the target ``separated'' (trained only on source domain) 
and ``mixed'' target (25\% merged from source domain). It can be seen that the BarrettsSegUDA model offers better performance in IoU and DSC than U-Net model by 5.7\% and 6.5\%, respectively. It can also be seen in Fig.~\ref{fig:qualitativeResults} (II) that the predicted masks using proposed UDA model for the post-acetic acid WL image is more similar to the ground truth images in comparison to those predicted by the U-Net model. 

\section{CONCLUSION}
We have applied an unsupervised domain adaptation framework for segmentation of Barrett's oesophagus area, an early cancer precursor. The proposed approach provided a generalised prediction of segmentation masks of unlabelled endoscopy images in cross modalities. The method does not depend on the existence of target labels and hence provided accurate and robust results in different imaging conditions. The experimental results demonstrated the effectiveness of the UDA model. In the future work, we will assess how the UDA framework can be extended to the semi-supervised domain adaptation setting. 

\small{\subsubsection*{Compliance with Ethical Standards}}
All endoscopy images used in this study were acquired from the videos that were consented for recording and analysis. The study was approved by the local Research Ethics Committee (REC Ref: 16/YH/0247).


\small{\subsubsection*{ACKNOWLEDGEMENTS}}
NC is funded by a private foundation, SG is funded by Boston Scientific, SA and JR are supported by the NIHR Oxford BRC. The authors would also like to acknowledge Prof. Barbara Braden and Dr. Adam Bailey for providing their clinical expertise. The authors declare no competing interests.

\bibliographystyle{IEEEbib}
\bibliography{refs}

\begin{thebibliography}{10}

\bibitem{cook2018cancer}
Michael~B Cook, Sally~B Coburn, Jameson~R Lam, Philip~R Taylor, Jennifer~L
  Schneider, and Douglas~A Corley,
\newblock ``Cancer incidence and mortality risks in a large us {Barrett's
  oesophagus} cohort,''
\newblock {\em Gut}, vol. 67, no. 3, pp. 418--529, 2018.

\bibitem{FCN}
J.~{Long}, E.~{Shelhamer}, and T.~{Darrell},
\newblock ``Fully convolutional networks for semantic segmentation,''
\newblock in {\em 2015 IEEE Conference on Computer Vision and Pattern
  Recognition (CVPR)}, 2015, pp. 3431--3440.

\bibitem{Olaf:MICCAI2015}
Olaf Ronneberger, Philipp Fischer, and Thomas Brox,
\newblock ``{U-Net: Convolutional Networks for Biomedical Image
  Segmentation},''
\newblock in {\em Medical Image Computing and Computer-Assisted Intervention --
  {MICCAI} 2015}, 2015, pp. 234--241.

\bibitem{DeepLabv3}
L.~{Chen}, G.~{Papandreou}, I.~{Kokkinos}, K.~{Murphy}, and A.~L. {Yuille},
\newblock ``Deeplab: Semantic image segmentation with deep convolutional nets,
  atrous convolution, and fully connected crfs,''
\newblock {\em IEEE Transactions on Pattern Analysis and Machine Intelligence},
  vol. 40, no. 4, pp. 834--848, 2018.

\bibitem{ali2020objective}
Sharib Ali, Felix Zhou, Barbara Braden, Adam Bailey, Suhui Yang, Guanju Cheng,
  Pengyi Zhang, Xiaoqiong Li, Maxime Kayser, Roger~D Soberanis-Mukul, et~al.,
\newblock ``An objective comparison of detection and segmentation algorithms
  for artefacts in clinical endoscopy,''
\newblock {\em Scientific Reports}, vol. 10, no. 1, pp. 1--15, 2020.

\bibitem{EndoCV2020}
Sharib Ali, Mariia Dmitrieva, Noha~M. Ghatwary, Sophia Bano, Gorkem Polat,
  Alptekin Temizel, Adrian Krenzer, Amar Hekalo, Yun~Bo Guo, Bogdan~J.
  Matuszewski, Mourad Gridach, Irina Voiculescu, Vishnusai Yoganand, Arnav
  Chavan, Aryan Raj, Nhan~T. Nguyen, Dat~Q. Tran, L{\^{e}}~Duy Huynh, Nicolas
  Boutry, Shahadate Rezvy, Haijian Chen, Yoon~Ho Choi, Anand Subramanian,
  Velmurugan Balasubramanian, Xiaohong~W. Gao, Hongyu Hu, Yusheng Liao, Danail
  Stoyanov, Christian Daul, Stefano Realdon, Renato Cannizzaro, Dominique
  Lamarque, Terry Tran{-}Nguyen, Adam Bailey, Barbara Braden, James East, and
  Jens Rittscher,
\newblock ``A translational pathway of deep learning methods in
  gastrointestinal endoscopy,''
\newblock {\em CoRR}, vol. abs/2010.06034, 2020.

\bibitem{guo2020polyp}
Yunbo Guo, Jorge Bernal, and Bogdan J~Matuszewski,
\newblock ``Polyp segmentation with fully convolutional deep neural
  networks—extended evaluation study,''
\newblock {\em Journal of Imaging}, vol. 6, no. 7, pp. 69, 2020.

\bibitem{Ali20203D}
Sharib Ali, Adam Bailey, James~Edward East, Simon~J. Leedham, Maryam Haghighat,
  TGU Investigators, Xin Lu, Jens Rittscher, and Barbara Braden,
\newblock ``Artificial intelligence-driven real-time {3D} surface
  quantification of barrett{\textquoteright}s oesophagus for risk
  stratification and therapeutic response monitoring,''
\newblock {\em medRxiv}, 2020.

\bibitem{gupta2020mi}
Soumya Gupta, Sharib Ali, Louise Goldsmith, Ben Turney, and Jens Rittscher,
\newblock ``{MI-UNet:} improved segmentation in ureteroscopy,''
\newblock in {\em 2020 IEEE 17th International Symposium on Biomedical Imaging
  (ISBI)}. IEEE, 2020, pp. 212--216.

\bibitem{ADDA}
E.~{Tzeng}, J.~{Hoffman}, K.~{Saenko}, and T.~{Darrell},
\newblock ``Adversarial discriminative domain adaptation,''
\newblock in {\em 2017 IEEE Conference on Computer Vision and Pattern
  Recognition (CVPR)}, 2017, pp. 2962--2971.

\bibitem{KDSS}
Y.~{Liu}, K.~{Chen}, C.~{Liu}, Z.~{Qin}, Z.~{Luo}, and J.~{Wang},
\newblock ``Structured knowledge distillation for semantic segmentation,''
\newblock in {\em 2019 IEEE/CVF Conference on Computer Vision and Pattern
  Recognition (CVPR)}, 2019, pp. 2599--2608.

\bibitem{Tsai}
Y.~{Tsai}, W.~{Hung}, S.~{Schulter}, K.~{Sohn}, M.~{Yang}, and M.~{Chandraker},
\newblock ``Learning to adapt structured output space for semantic
  segmentation,''
\newblock in {\em 2018 IEEE/CVF Conference on Computer Vision and Pattern
  Recognition}, 2018, pp. 7472--7481.

\bibitem{CardioUDA}
Nanqing Dong, Michael Kampffmeyer, Xiaodan Liang, Zeya Wang, Wei Dai, and Eric
  Xing,
\newblock ``Unsupervised domain adaptation for automatic estimation of
  cardiothoracic ratio,''
\newblock in {\em Medical Image Computing and Computer Assisted Intervention --
  {MICCAI 2018}}. 2018, pp. 544--552, Springer International Publishing.

\bibitem{haq20a}
Mohammad~Minhazul Haq and Junzhou Huang,
\newblock ``Adversarial domain adaptation for cell segmentation,''
\newblock Montreal, QC, Canada, 06--08 Jul 2020, vol. 121 of {\em Proceedings
  of Machine Learning Research}, pp. 277--287, PMLR.

\bibitem{pandey2020unsupervised}
Prashant Pandey, Aayush~Kumar Tyagi, Sameer Ambekar, and Prathosh AP,
\newblock ``{Unsupervised Domain Adaptation for Semantic Segmentation of {NIR}
  Images through Generative Latent Search},''
\newblock in {\em European conference on computer vision {(ECCV)}}, 2020.

\bibitem{tan2020efficientnet}
Mingxing Tan and Quoc~V. Le,
\newblock ``Efficientnet: Rethinking model scaling for convolutional neural
  networks,''
\newblock {\em CoRR}, vol. abs/1905.11946, 2019.

\end{thebibliography}
\end{document}